# Stable ion-tunable antiambipolarity in mixed ion-electron conducting polymers enables biorealistic artificial neurons

Padinhare Cholakkal Harikesh[1,†], Chi-Yuan Yang[1,†], Han-Yan Wu[1], Silan Zhang[1,2], Jun-Da Huang[1], Magnus Berggren[1,2,3], Deyu Tu[1,†], Simone Fabiano[1,2,3*]

[1]Laboratory of Organic Electronics, Department of Science and Technology, Linköping University, SE-601 74 Norrköping, Sweden.

[2]Wallenberg Wood Science Center, Linköping University, SE-601 74 Norrköping, Sweden.

[3]n-Ink AB, Teknikringen 7, 583 30 Linköping, Sweden.

[†]Contributed equally

Correspondence should be addressed to: simone.fabiano@liu.se

**Bio-integrated neuromorphic systems promise for new protocols to record and regulate the signaling of biological systems. Making such artificial neural circuits successful requires minimal circuit complexity and ion-based operating mechanisms similar to that of biology. However, simple leaky integrate-and-fire model neurons, commonly realized in either silicon[1] or organic[2,3] semiconductor neuromorphic systems, can emulate only a few neural features. More functional neuron models[1,4,5], based on traditional complex Si-based complementary-metal-oxide-semiconductor (CMOS)[6–8] or negative differential resistance (NDR)[9–11] device circuits, are complicated to fabricate, not biocompatible, and lack ion- and chemical-based modulation features. Here we report a biorealistic conductance-based organic electrochemical neuron (c-OECN) using a mixed ion-electron conducting ladder-type polymer with reliable ion-tunable antiambipolarity. The latter is used to emulate the activation/inactivation of Na channels and delayed activation of K channels of biological neurons. These c-OECNs can then spike at bioplausible frequencies nearing 100 Hz, emulate most critical biological neural features, demonstrate stochastic spiking, and enable neurotransmitter and $Ca^{2+}$-based spiking modulation. These combined features are impossible to achieve using previous technologies.**

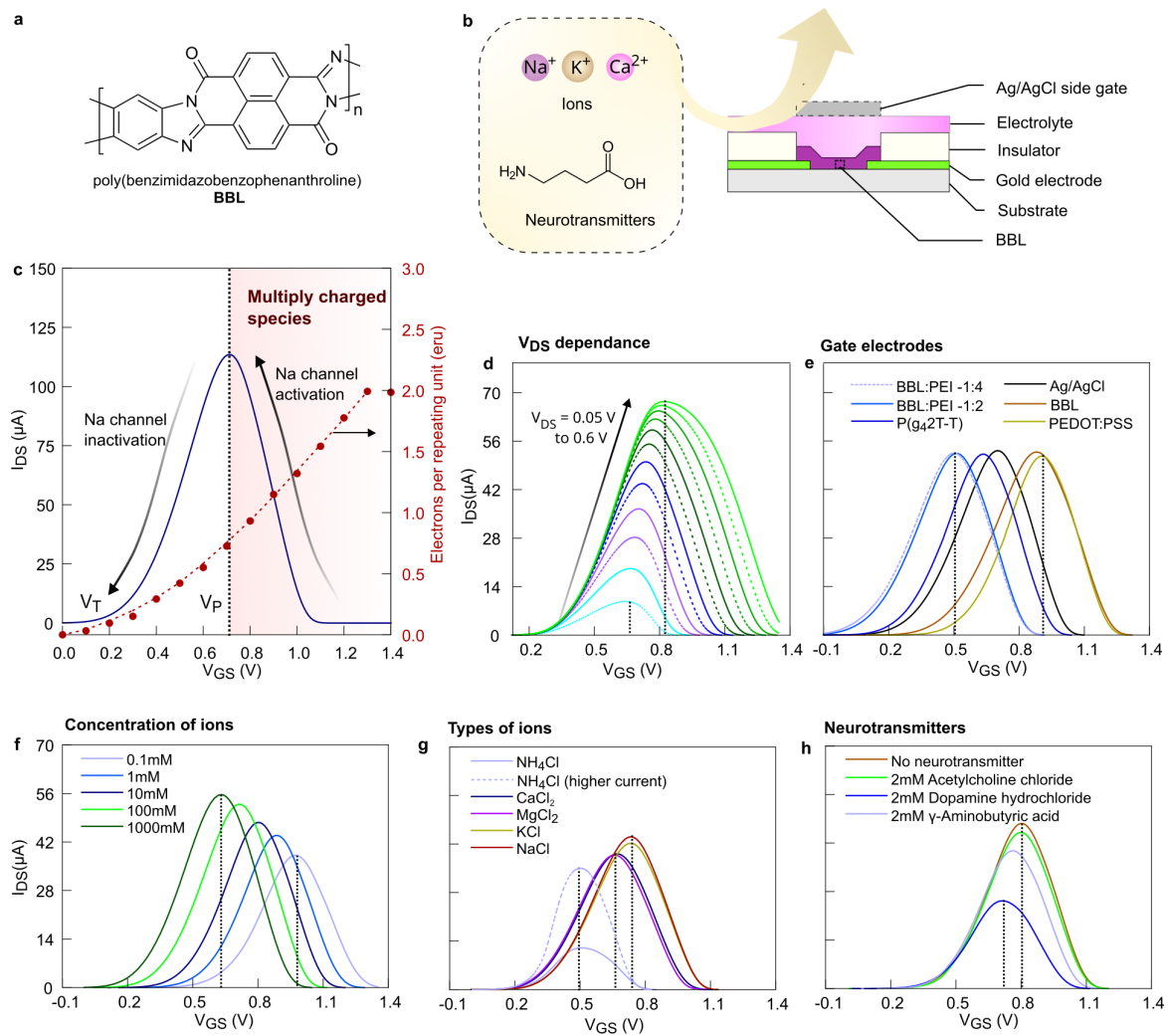

**Figure 1**: (a, b) The structure of BBL and the OECT device. (c) The antiambipolar behaviour in BBL resembling the activation and inactivation of Na channels in a neuron. (d-h) Modulation of the antiambipolar behaviour using electrical and chemical means. The OECT used in the comparison has a $W/L = 40$ μm/6 μm and 20 nm thick BBL except for the higher current NH$_4$Cl device, which uses a wider channel ($W/L = 400$ μm/6 μm). A concentration of 100 mM is used for comparing various types of ions. Neurotransmitter studies are carried out in 10 mM NaCl.

Neurons communicate and process information using action potentials or spikes in membrane potentials[12]. This spike generation and the properties of the neuron are defined by voltage-, ion-, and neurotransmitter-dependent conductance of various ion channels in the cell membrane. On receiving an electrical input, the cumulative effect of ionic currents through these channels–primarily the sodium and potassium channels–perturbs the membrane voltage from its resting value, resulting in an action potential. Thus, to create a realistic electrical circuit analogue to the biological neuron, one should emulate the conductance of the sodium channel, which shows a fast activation and inactivation, and the potassium channel, which activates after a certain delay. Emulating such a conductance-based model using Si CMOS circuits requires a large number of transistors (close to 15-30)[6,13]. Alternative technologies like Mott-memristor-based NDR devices[9,10] or antiambipolar p-n heterojunction[11] transistors can approximate realistic neuron behavior with fewer components. However, they are not biocompatible and are also restricted to pure electrical modulation of neural features without exploring ion/neurotransmitter-based modulation mechanisms of real biological neurons. Organic mixed ion-electron conducting polymers are interesting in this context due to their biocompatibility and coupled ionic-electronic transport properties that are amenable to modulation by external dopants[14,15]. Recently, we observed that rigid conjugated polymers like the ladder-type poly(benzimidazobenzophenanthroline) (BBL) exhibit a reduction in the electrical conductivity on high electrochemical doping (>0.8 electron/monomer) due to the formation of multiply charged species with reduced mobility[16] (**Figure 1c**). When used as the channel material in organic electrochemical transistors (OECTs), BBL exhibits a unique, stable, and reversible Gaussian-shaped transfer curve (or antiambipolar behavior) which is similar to voltage-controlled NDR but here instead realized in a three-terminal configuration (**Figure 1a-b, Supplementary Figure 1-4**). When implemented in a circuit, the two sides of this gaussian current evolution can be analogous to the activated and inactivated states of the voltage-gated Na channel in the Hodgkin-Huxley neuron model (HH model, **Supplementary Note 1**)[4]. Although such antiambipolar behavior can also be observed in other n- and p-type polymers (**Supplementary Figure 5**), only BBL with suitable electron affinity (4.15 eV) and a rigid ladder-like structure composed of double-strand chains linked by condensed π-conjugated units can sustain such high doping levels without any conformational disorder[17], enabling reversibility.

The OECT configuration provides improved control over the antiambipolar response compared to a conventional two-terminal NDR device. For example, applying a higher drain voltage ($V_{DS}$) increases the peak current, causes a shift in the voltage of the peak current ($V_P$), and results in a gaussian distribution of a greater full-width at half maximum (FWHM) (**Figure 1d**) due to variable doping levels at the drain and source electrodes. The $V_P$ and threshold voltage ($V_T$) can also be shifted and controlled by using gate electrodes of appropriate work function (**Figure 1e**) and by tuning the concentration of ions in the electrolyte (**Figure 1f**). In addition, for a given concentration (100 mM), different ions shift the transfer curve by varying degrees, with $Ca^{2+}$, $Mg^{2+}$, and $NH_4^+$ showing lower $V_P$ as compared to $Na^+$ and $K^+$ **(Figure 1g)**. Interestingly, various ammonium-based organic cations also lead to unique gaussian behaviors **(Supplementary Figure 6)**, unlocking the possibility of chemical-specific responses. Hence it is possible to tune the antiambipolar response using different neurotransmitters like acetylcholine, dopamine, and γ-butyric acid (GABA) with different configurations of amine groups **(Figure 1h)**. Adding 2 mM acetylcholine to the 100 mM NaCl electrolyte of the OECT does not shift the $V_P$, whereas the same concentration of dopamine and GABA cause changes in $V_P$ and the channel's conductance. We hypothesize that this effect, which is reversible on removing the neurotransmitter, is due to differences in interactions between these molecules and BBL and changes in pH.

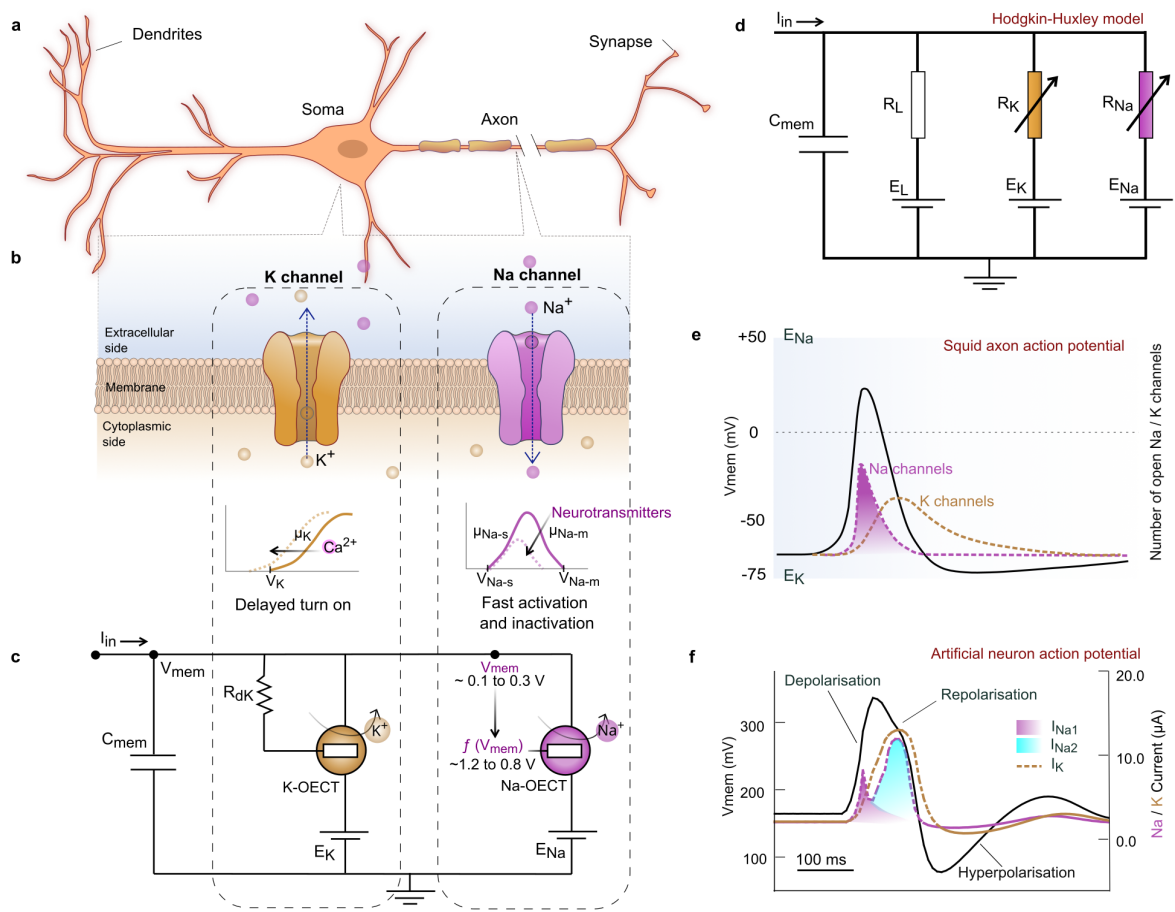

**Figure 2 : (a-c)** The analogy between the biological neuron showing the Na and K ion channels and the c-OECN circuit with Na and K ion based OECTs and their modulation with Ca and Neurotransmitters. $C_{mem}$ is optional in the circuit and can be embedded in the intrinsic

capacitance of the OECT **(d)** The Hodgkin-Huxley circuit of neuron. The comparison of **(e)** Squid axon action potential and **(f)** OECN action potential.

The c-OECN circuit described here uses two OECTs – one $Na^+$ ion-based (Na-OECT) and another $K^+$ ion-based (K-OECT) coupled to two voltage sources $E_{Na}$ (500 mV) and $E_K$ (~ 12 to -72 mV) like the two ion channels and batteries in the HH model[4,18] **(Figure 2 a-d)**. The switching speed of these OECTs is in the range of 0.5-1 ms[19], which is comparable to the time scales of sodium and potassium channel activation in biological neurons. The K-OECT has a thicker BBL film (50 nm compared to 20 nm for the Na-OECT) to allow higher currents through the potassium channel. A current $I_{in}$ injected into this circuit is integrated by the capacitance $C_{mem}$ causing the voltage $V_{mem}$ to increase from its resting value of ~175 mV. This simultaneously sweeps the voltage on the gate of Na-OECT from ~1.2 V to around 0.8 V using an inverting amplifier **(Supplementary Figure 7)** to traverse the antiambipolar transfer curve through $V_P$, producing a spike in current $I_{Na1}$. This current spike charges the capacitor even further, resulting in a rapid increase of $V_{mem}$ to ~0.34 V (Depolarisation). The K-OECT, which turns on after a delay enabled by its intrinsic gate capacitance and resistance $R_{dk}$, reaches its peak current after Na-OECT has crossed its maximum at $V_P$. The capacitor is thus discharged through K-OECT, causing the $V_{mem}$ to drop (Repolarisation) and traversing the gaussian response to its initial state back through $V_P$, causing a secondary spike $I_{Na2}$. The secondary spike in Na-OECT (not present in squid axon action potential, **Figure 2 e,f**) is unnecessary for spike generation but is unavoidable; hence is entirely discharged by the K-OECT not to cause any further voltage increase. Since the current of K-OECT is higher and persists longer, the voltage is brought below the resting value of 175 mV for a brief period (Hyperpolarisation). All these processes repeat cyclically if the input current remains constant, resulting in continuous action potential generation (Tonic spiking). The c-OECN action potential thus shows typical features of a biological action potential, including depolarisation, repolarisation, and hyperpolarisation (**Figure 2f**). **Supplementary Note 2** provides the circuit analysis of the c-OECN and the SPICE simulation and a comparison of the circuit equations with the HH model.

The resemblance of the operation of c-OECN with the biological neuron enables mimicking several of the neural features[5] by modulating the threshold and currents of OECTs **(Figure 3, Supplementary Table 1).** In the standard configuration, the neuron exhibits tonic spiking, i.e. excitability in the presence of a constant input while remaining quiescent otherwise. Tuning the capacitance $C_{mem}$ and the resistance $R_{dk}$ can tune the frequency of this spiking. The c-OECN spikes at a frequency of around 5 Hz with a $C_{mem}$ = 1 μF and $R_{dk}$ = 470 KΩ (**Figure 3a**) and can be increased to reach 80 Hz (**Figure 3c**) by excluding external capacitance and then utilizing only the internal capacitance of the OECT. Hence, biorealistic frequencies can be achieved with a neuron based on only three transistors and two resistors, which to the best of our knowledge, is the lowest number of components for any conductance-based neuron. For the ease of measurement of neural features with various pulsed inputs, we used the lower frequency c-OECN (~5 Hz) by employing a $C_{mem}$ of 1 μF.

Most neural features like tonic spiking, latency, subthreshold oscillations, integration, refractoriness, resonance, threshold variability, rebound spiking, accommodation, phasic spiking, phasic bursting, class 1 and class 2 excitability can be demonstrated using this circuit. Each of these has a specific function or serves as a mathematical operator in a neuron (**Supplementary Note 4**). Switching from class 1 (input strength dependant excitability) to class 2 (spiking only at high current inputs with a high frequency) and class 3 spiking (spiking/bursting only at the beginning of input) can be enabled by simple tuning of the threshold of the K-OECT (by changing $E_k$) to modify the relative timing of it turning on. For example, for the c-OECN demonstrated here, an $E_k$ of -50 mV results in a class 1 spiking neuron, while causing K-OECT to turn on earlier by changing $E_k$ to -65 mV changes the behavior to class 2 spiking and by increasing it to a more negative value of -70 mV (-72 mV) results in Phasic bursting (Phasic spiking) which is class 3 behavior.

The c-OECN can also exhibit input noise-dependent stochasticity or spike skipping, like the biological neurons[20]. The class 2 neuron shown in **Figure 3n** does not spike at an input current of 2 μA. However, when very low noise is superimposed on this input keeping the average current the same, it starts spiking at a particular frequency—increasing the input noise results in random skipping of spikes while keeping this base frequency constant. A similar mechanism is observed in mammalian cold thermoreceptors[20] where temperature-induced noise causes stochasticity in spiking and is used to extend the range of encodable stimuli. Such stochastic spiking can also enable probabilistic neural sampling and find application in spike-based Bayesian learning and inference[21].

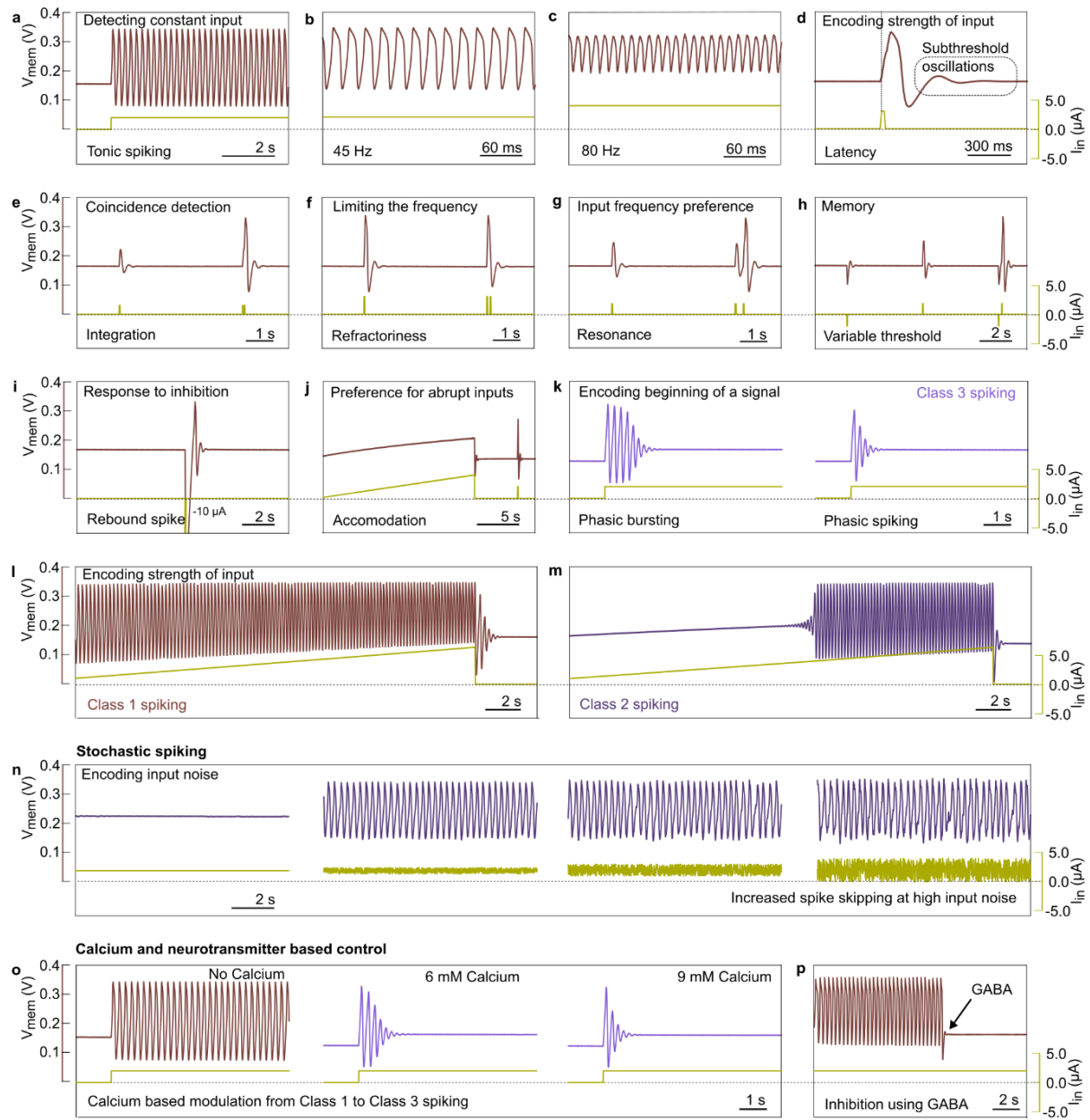

**Figure 3**: Experimental demonstration of various neural features using c-OECN and their functions in biology. (a) Tonic spiking, (b and c) higher frequency tonic spiking by modulating $R_{dk}$ and $C_{mem}$, (d) latency, (e) integration, (f) refractoriness, (g) resonance, (h) threshold variability, (i) rebound spike, (j) accommodation, (k) class 3 spiking- phasic bursting and spiking, (l) class 1 spiking, (m) class 2 spiking, (n) stochastic spiking with noisy input, (o) calcium-based modulation from class 1 to class 3 spiking and (p) inhibition of spiking using γ-aminobutyric acid. The parameters used to enable these behaviours are listed in Supplementary Table 1.

The unique feature of the c-OECN compared to other conductance-based circuit realizations of neurons is that it can be controlled using secondary ions like $Ca^{2+}$ and neurotransmitters as they can affect the $V_P$ and maximum current of the gaussian transfer curve. In biological neurons, $Ca^{2+}$ has crucial roles in regulating neural activity by modifying the opening and closing of Na and K channels, stimulating the release of neurotransmitters leading to synaptic plasticity, and even in the regulation of metabolism and cell growth[22]. Inspired by this, we tried to modulate the K channel of the c-OECN using external $Ca^{2+}$ ions. The incorporation of $Ca^{2+}$ ions into the electrolyte of the K-OECT shifts its threshold $V_K$ towards lower values. This is equivalent to $V_K$ modulation by altering $E_K$ as described above and hence creates the same effect, i.e., a shift from class 1 spiking to phasic bursting and finally phasic spiking on a slow increase of $Ca^{2+}$ concentration (**Figure 3o**). Such a transition is similar to the case in biological neurons, where the generation of phasic firing is known to be $Ca^{2+}$ ion concentration dependent[23]. In addition to ions, biological neurons are also affected by the presence of neurotransmitters. For example, γ-aminobutyric acid (GABA) is an inhibitory neurotransmitter in the brain[24] and inhibits the generation of action potentials by increasing chloride or potassium ion conductance and hence hyperpolarizing the membrane. Here we coupled GABA to the Na-OECT, and since GABA reduces the maximum current of the Na-OECT and shifts its $V_P$ towards lower values, the Na current spike induced depolarization, and hence spiking is instantly inhibited in the c-OECN, thus enabling neurotransmitter-induced modulation of spiking similar to biology.

**Table 1**: Comparison of artificial neurons based on various technologies

|  | Low circuit elements | Neural features | Biorealistic switching speeds | Ion based modulation | Biocompatibility | Low voltage operation |
|---|---|---|---|---|---|---|
| Silicon CMOS (biorealistic models)[1,6,7,13,25–27] | ● | ●●●● | ●●● | ○ | ●● | ●● |
| Mott-Memristors[9,10] | ●●● | ●●●● | ●●● | ○ | ○ | ●●● |
| 2D material-based Gaussian heterojunction[11] | ●●● | ●● | ● | ○ | ○ | ●● |
| OFET (LIF)[3] | ●●● | ● | ●● | ○ | ●●● | ●● |
| Complementary OECT (LIF)[2] | ●●● | ● | ● | ●● | ●●●● | ●●●● |
| **Antiambipolar-OECT (this work)** | ●●●● | ●●● | ●●●● | ●●●● | ●●●● | ●●●● |

We demonstrated a biorealistic OECN based on highly tunable, stable, and reversible antiambipolar behaviour in BBL-based OECTs. Two OECTs modulated by Na$^+$ and K$^+$ ions resemble the voltage-gated ion channels in biological neurons, enabling various neural features with only three transistors, two resistors, and a capacitor (optional). A comparison of the c-OECN with other neuron technologies is provided in **Table 1**. The intrinsic capacitance of the OECTs and the biorealistic switching speeds circumvent the need for additional capacitors required in Si-based circuits to spike at biologically plausible frequencies for interaction with real-world events and biological neurons. Alternative implementations using Mott-memristors that can exhibit similar features are inherently faster than biology (ns to μs) and are associated with unfavourable temperature increases on operation, making them unsuitable for biointegration. In addition, using single polymer material to achieve NDR like gaussian behaviour averts complex fabrication strategies and helps achieve lower device dimensions compared to the heterojunction approach in 2D materials or p-n junction polymers. The ion and neurotransmitter-based modulation in this c-OECN is unprecedented and holds the possibility to be extended to other biomolecules which can interact with BBL. Similar stable ladder-like conjugated polymers functionalized to interact with specific biomolecules is a possible way forward for realizing intelligent closed-loop sensing and feedback neuromorphic systems and future brain-machine interfaces.

**Methods:**

**Materials** : BBL is synthesized following a procedure reported previously ($\eta$ = 6.3 dL g$^{-1}$ in MSA at 30 °C, $M_w$ = 35 kDa)[19,28]. PEDOT:PSS (Clevios PH1000) is purchased from Heraeus Holding GmbH. MSA, chloroform, 1,2-dichlorobenzene, ethylene glycol, (3-Glycidyloxypropyl)trimethoxysilane, and 4-dodecylbenzenesulfonic acid are purchased from Sigma-Aldrich.

**Thin-Film Casting** : BBL is dissolved in MSA at 100 °C for 12 h and cooled to room temperature to obtain a BBL-MSA solution. This solution is spin-coated (1000 rpm, 60 s, acceleration 1000 rpm s$^{-1}$) on OECT substrates. Then, the thin BBL films are immersed in deionized water to remove residual MSA and dried with nitrogen flow. PEDOT:PSS, ethylene glycol, (3-glycidyloxypropyl)trimethoxysilane, and 4-dodecylbenzenesulfonic acid are mixed in the volume ratio of 100 : 5 : 1 : 0.1 and sonicated for 10 min. This mixed solution is spin-coated (2000 rpm, 60 s, acceleration 2000 rpm s$^{-1}$) on OECT substrates and annealed at 120 °C for 1 min to crosslink PEDOT:PSS.

**OECT fabrication and testing** : OECTs are fabricated following a procedure reported previously[19,29]. 4-inch glass wafers are cleaned via successive sonication in acetone, deionized water, and isopropyl alcohol, and then dried with nitrogen. Source/drain electrodes (5 nm Cr and 50 nm Au) are thermally deposited and photolithographically patterned by wet etching. A first layer of parylene C (1 µm), deposited together with a small amount of 3-(trimethoxysilyl)propyl methacrylate (A-174 Silane) to enhance adhesion, acts as an insulator to prevent disturbing capacitive effects at the metal liquid interface. Subsequently, a dilution of industrial surfactant (2% Micro-90) is spin-coated as an antiadhesive layer and a second layer of parylene C (2 µm) is deposited as a sacrificial layer. To protect the parylene C layers from a subsequent plasma reactive ion etching step (150 W, $O_2$ = 500 sccm, $CF_4$ = 100 sccm, 380 s), a thick positive photoresist (5 µm, AZ10XT520CP) is spin-coated on the parylene C layer. A second photolithographic patterning step is carried out to define the contact pads and the OECT channel, and the AZ developer is applied to the photoresist. The subsequent plasma reactive ion etching step indiscriminately removes the layer of organic materials, including both photoresist and parylene C, so that the contact pads and the OECT channel area are exposed to the air while other areas are still covered with two layers of parylene C. The channel between source and drain are patterned to obtain $W/L$ = 40 µm/6 µm or 80 µm/6 µm (for Na-OECTs), and 400 µm/6 µm (for K-OECTs), BBL-MSA solutions are spin-coated to obtain 20 nm thick (Na-OECT) or 50 nm thick (K-OECT) film, covering the whole substrate surface. The sacrificial layer is peeled off and the BBL film on it is removed, leaving separated pieces of film staying in the wells, consisting of the semiconductor connecting the OECT source/drain electrodes. Ag/AgCl paste is drop-casted on the substrate to form 1 µm thick, 9 mm$^2$ square gate electrode. PEDOT:PSS, p($g_7NC_{10}N$) and P($g_42T-T$)-based OECTs are fabricated using a similar procedure. For all OECTs, 0.1 M NaCl aqueous solution is used as the electrolyte unless specified otherwise. The OECTs are characterized using Keithley 4200A-SCS.

**SPICE Simulation**: The SPICE models of K-OECT and Na-OECT are developed in B2 SPICE (EMAG Technologies Inc.) Both OECT models are built in order to simulate the spiking features of c-OECNs. The details of the simulation are presented in **Supplementary Note 4**.

**Acknowledgements**

We acknowledge Renee Kroon for synthesizing P(g$_4$2T-T) and Iain McCulloch for providing p(g$_7$NC$_{10}$N) that are used to compare antiambipolar behavior. We thank Kai Xu for discussing antiambipolar behavior in BBL. This work was financially supported by the Knut and Alice Wallenberg Foundation, the Swedish Research Council (2020-03243), Olle Engkvists Stiftelse (204-0256), VINNOVA (2020-05223), the European Research Council (834677 "e-NeuroPharma" ERC-2018-ADG), the European Commission through the FET-OPEN project MITICS (GA-964677), and the Swedish Government Strategic Research Area in Materials Science on Functional Materials at Linköping University (Faculty Grant SFO-Mat-LiU 2009-00971).

**Author contributions**

P.C.H, D.T, and S.F. conceived and designed the experiments. P.C.H and D.T fabricated and tested the circuits and analyzed the data. C.-Y.Y and H.-Y.W performed the synthesis of materials, fabrication, and characterization of the OECTs for the circuit. C.-Y.Y and J.-D.H compared the antiambipolar behavior in various polymers. S.Z. studied the effect of ions on antiambipolar behavior. D.T. designed the circuit and conducted the simulations. P.C.H, D.T, M.B, and S.F wrote the manuscript. All authors contributed to the discussion and manuscript preparation.

**Competing interests**

The authors declare no competing interests.

**Additional information**

**Correspondence and requests for materials** should be addressed to S.F.

# Supplementary Information

**Stable ion-tunable antiambipolarity in mixed ion-electron conducting polymers enables biorealistic artificial neurons**

Padinhare Cholakkal Harikesh[1,†], Chi-Yuan Yang[1,†], Han-Yan Wu[1], Silan Zhang[1,2], Jun-Da Huang[1], Magnus Berggren[1,2,3], Deyu Tu[1,†], Simone Fabiano[1,2,3*]

[1]Laboratory of Organic Electronics, Department of Science and Technology, Linköping University, SE-601 74 Norrköping, Sweden.

[2]Wallenberg Wood Science Center, Linköping University, SE-601 74 Norrköping, Sweden.

[3]n-Ink AB, Teknikringen 7, 583 30 Linköping, Sweden.

[†]Contributed equally

Correspondence should be addressed to: simone.fabiano@liu.se

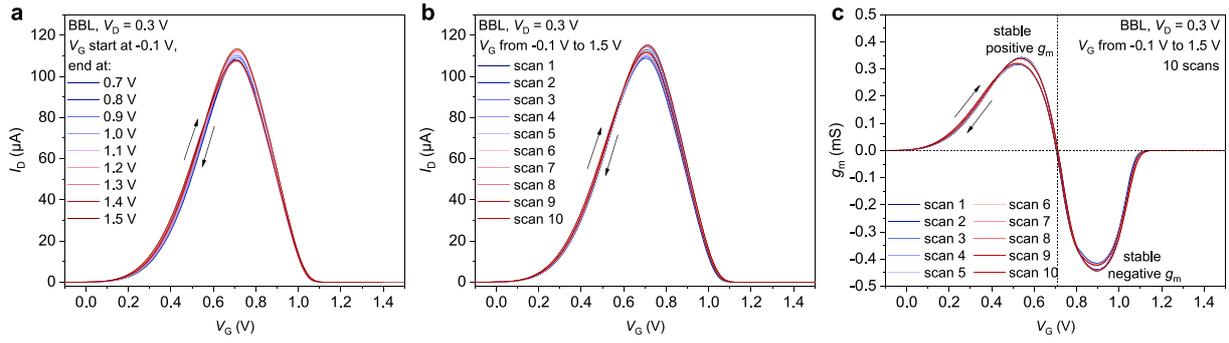

**Supplementary Figure 1:** (a and b) stable antiambipolar behaviour in BBL showing (c) positive and negative transconductance regions.

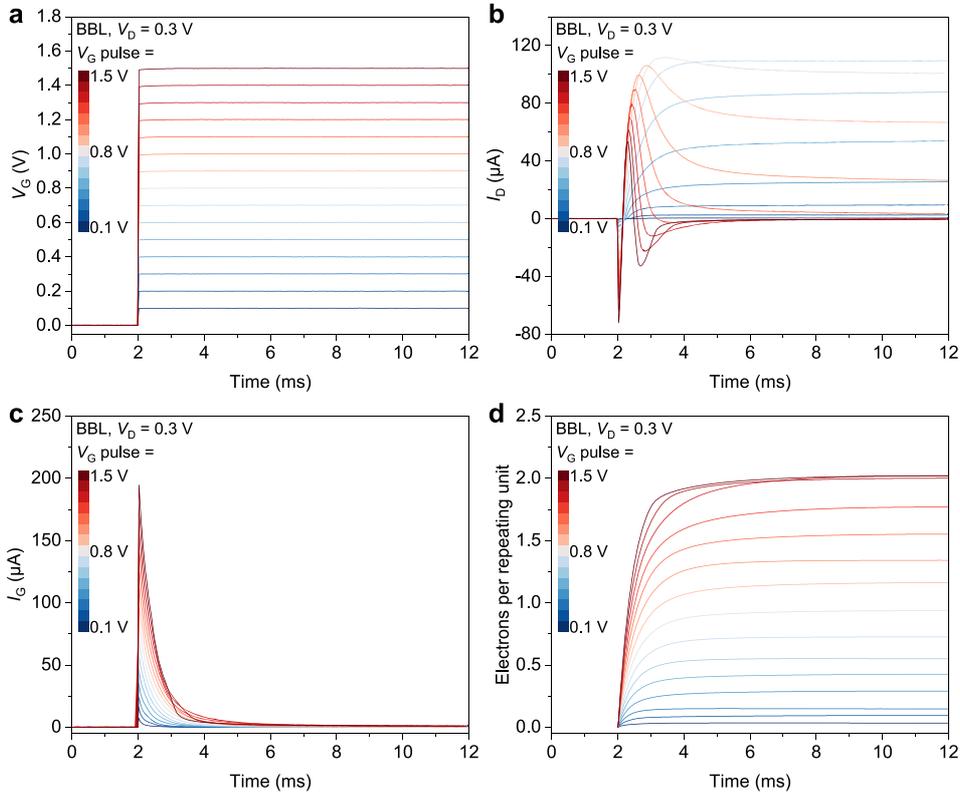

**Supplementary Figure 2**: Calculation of number of electrons injected per repeating unit into BBL during doping. (a) Applied gate voltage over 12 ms, (b and c) the corresponding drain and gate currents. (d) The electrons per repeating unit (eru), obtained by integrating the gate current over time using the relation :

$$eru = \frac{\int_{t_1}^{t_2} I_G \, dt}{F} \cdot \frac{M_{mol}}{\rho V}$$

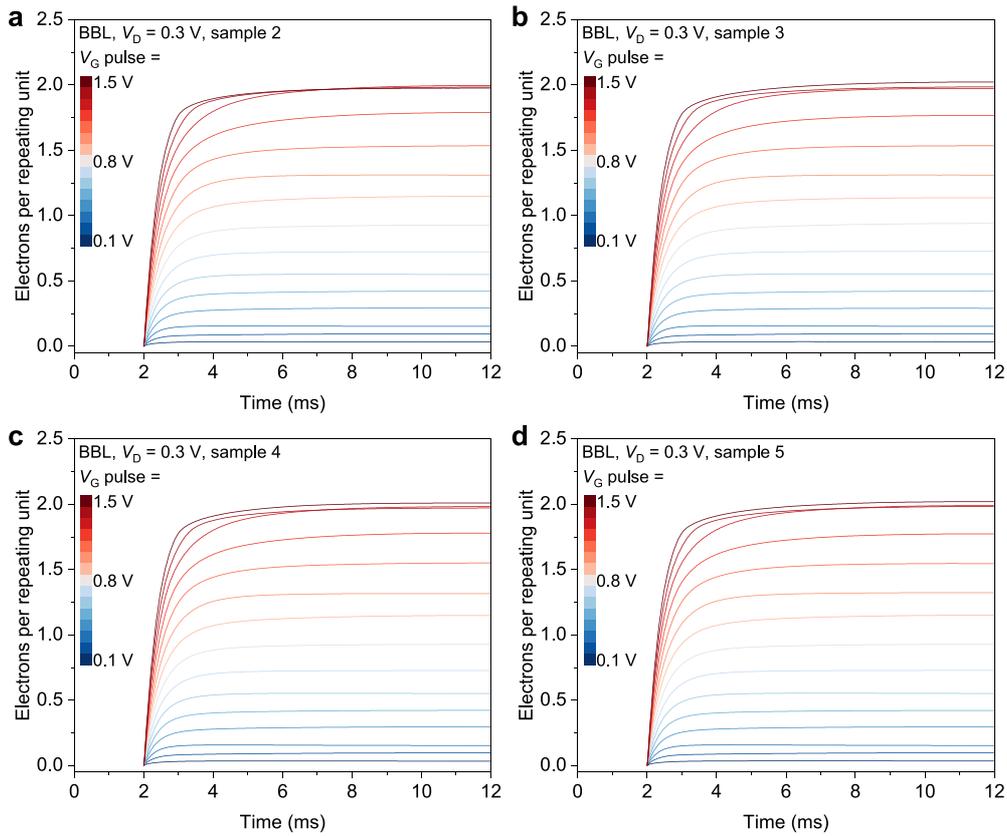

**Supplementary Figure 3 :** (a-d) Reproducibility of calculation of electrons injected per repeating unit across four samples

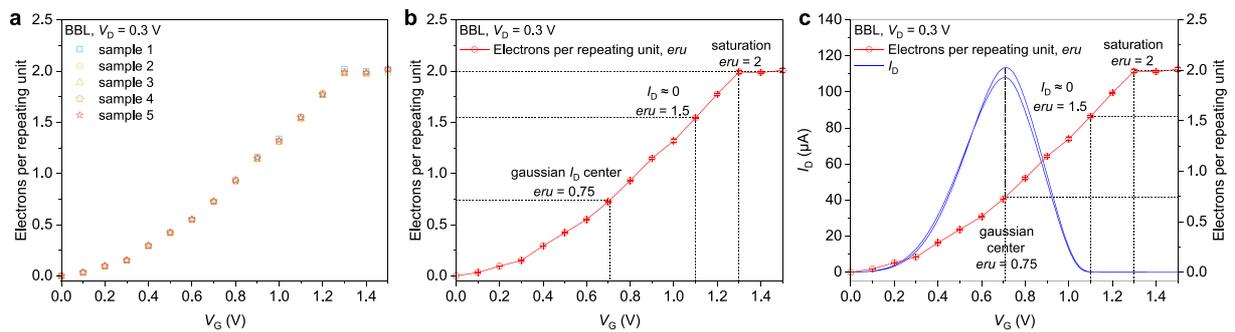

**Supplementary Figure 4:** (a-c) Correlation between applied gate voltage and electrons per repeating unit (eru) at a drain voltage of 0.3 V and the reproducibility across various samples.

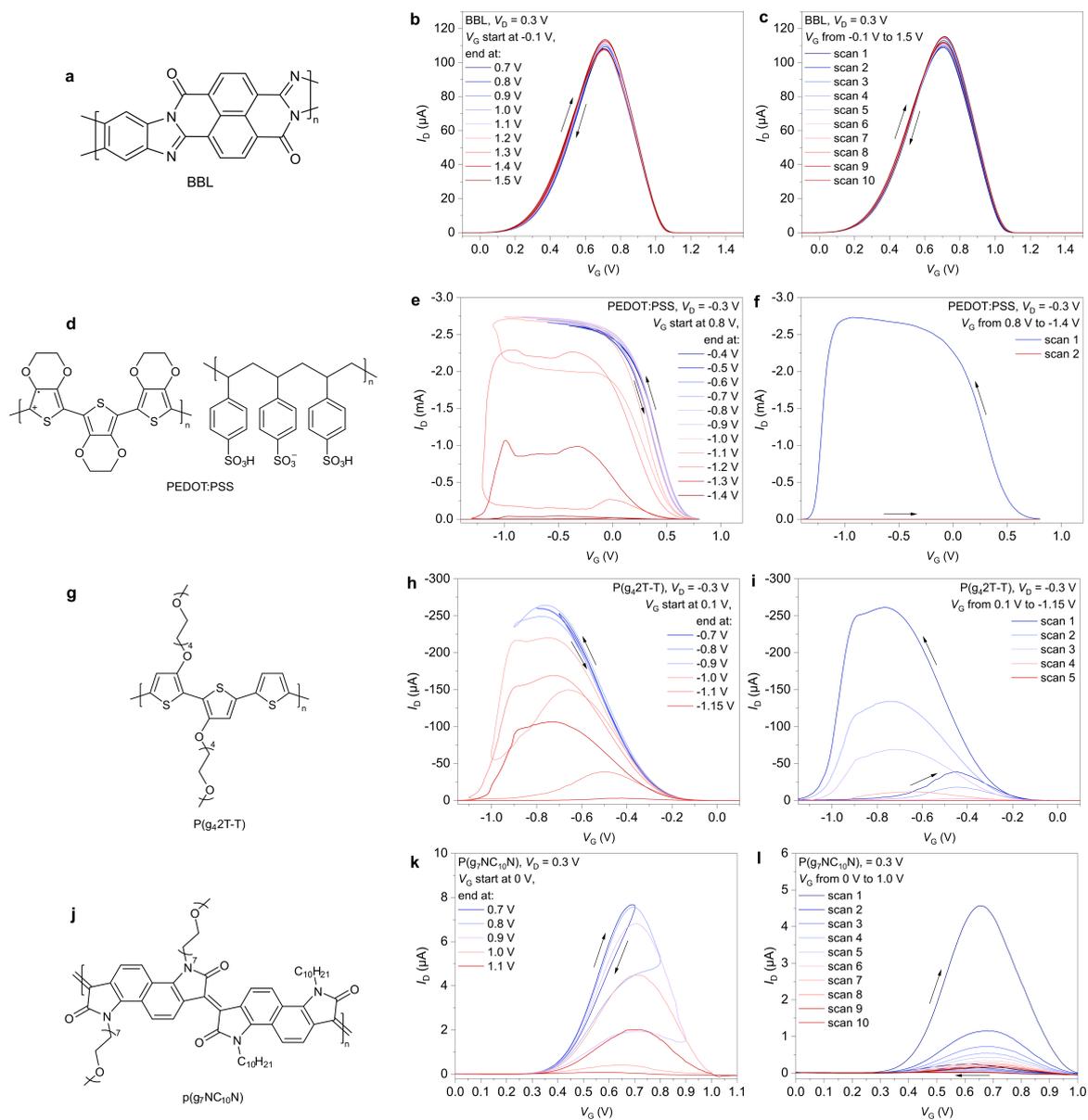

**Supplementary Figure 5 :** (a-c) The structure of BBL and the stability of antiambipolar behaviour. (d-k) Structures of PEDOT:PSS, P($g_4$2T-T), P($g_7$N$C_{10}$N) and the instability of antiambipolar behaviour in these polymers. P($g_4$2T-T) is dissolved in 1,2-dichlorobenzene and p($g_7$N$C_{10}$N) in chloroform and spin-coated (2000 rpm, 60 s, acceleration 2000 rpm s$^{-1}$) on OECT substrates.

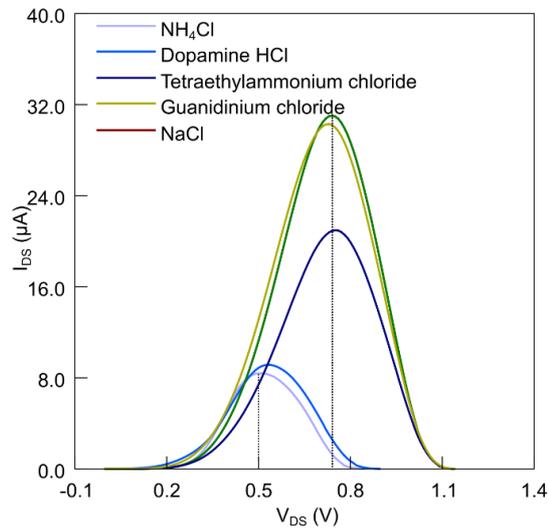

**Supplementary Figure 6 :** Effect of different ammonium cations on the antiambipolar behaviour in BBL.

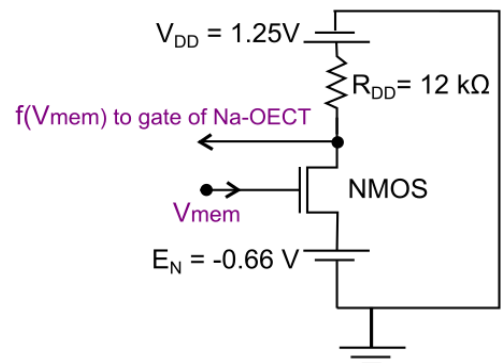

**Supplementary Figure 7 :** Inverting amplifier with a NMOS transistor (Infineon BSP295) and a load resistor ($R_{DD}$).

**Supplementary Note 1: Hodgkin-Huxley Neuron model**

The spiking in a neuron is caused by the opening and closing of voltage-gated Na and K ion channels. There is a concentration difference of ions inside and outside neuron maintained by ion pumps. This is modelled as Na and K batteries and form the lower and upper limits of the action potential. When a current is injected, the membrane capacitance $C_{mem}$ causes voltage to increase, and the voltage-gated Na channels activate and inactivate, followed by a delayed activation of K channel. The whole system can be modelled using four differential equations given by Hodgkin and Huxley.

In the circuit of the Hodgkin-Huxley model shown Figure 2d, the relation between neuron membrane voltage $V_{mem}$ and the influx of current $I_{in}(t)$ can be written as[2]:

$$I_{in}(t) = C_{mem}\frac{dV_{mem}}{dt} - \sum_k I_k(t), \tag{1}$$

where $C_{mem}$ is the membrane capacitance and $\sum_k I_k(t)$ is the sum of the ionic currents passing through the membrane in biological terms. Hodgkin and Huxley further formulated the three ionic currents of K channel, Na channel, and the leakage as

$$\sum_k I_k(t) = g_K n^4 (V_{mem} - E_K) + g_{Na} m^3 h (V_{mem} - E_{Na}) + g_L(V_{mem} - E_L), \tag{2}$$

where $E_K$, $E_{Na}$, and $E_L$ are the reversal potentials and *n*, *m*, and *h* are ion channel gating variables which control the activation of K channel, the activation of Na channel and the inactivation of Na channel, respectively.

**Supplementary Note 2 : OECT HH neuron circuit analysis and SPICE simulations**

The conductance of the K-OECT has a sigmoidal behaviour dependent on $V_{mem}$ based on its mobility $\mu_k$, threshold voltage $V_K$, and time (delay) defined by $R_{dk}$ and intrinsic capacitance $C_{dk}$ analogous to the K activation variable n in HH model. For the Na-OECT, the conductance has a gaussian behaviour defined by two threshold voltages ($V_{Na-m}$, $V_{Na-s}$) and mobilities ($\mu_{Na-m}$, $\mu_{Na-s}$) corresponding to the multiply and singly charged species on either side of the gaussian similar to the activation (m) and inactivation (h) variables of Na in HH model. $E_k$ and $E_{Na}$ are constants and act as sodium and potassium batteries defining the lower and upper bounds of action potential; $E_k$ also helps to modulate the threshold $V_k$ of K-OECT. $\mu_k$, $V_{Na-m}$, $\mu_{Na-m}$, $V_{Na-s}$ and $\mu_{Na-s}$ are functions of $V_{mem}$, while $V_{mem}$ is a function of time defined by the charging of the capacitance $C_{mem}$. Hence, the properties of the c-OECN can be described by these coupled state variables.

As shown in **Figure 2c**, the Hodgkin-Huxley neuron circuit based on BBL OECTs addresses the ionic channel (K and Na) conductance and their activation and inactivation, whereas the leakage channel is simply left to the device leakage in the circuit. The HH neuron model in Eq. 1 can be rewritten as follows

$$I_{in}(t) = C_{mem}\frac{dV_{mem}}{dt} + i_K(t) - i_{Na}(t), \qquad (3)$$

where $i_K(t)$ and $i_{Na}(t)$ are the ionic currents passing through the K channel and the Na channel respectively.

Given the K-OECT operates in a linear region when activated, the current of K channel is therefore expressed as

$$i_K(t) = \mu_K C_V \frac{Wd}{L}\left[\left(V_{mem}\left(1 - e^{\frac{t}{R_{dk}C_{dk}}}\right) - E_K - V_K\right)(V_{mem} - E_K) - \frac{(V_{mem}-E_K)^2}{2}\right], \qquad (4)$$

where $C_V$ is the volumetric capacitance of BBL polymer semiconductor, W, L, and d are the channel width, length, and thickness respectively, $R_{dk}$ and $C_{dk}$ provides the activation delay of the K-OECT. In the absence of $C_{dk}$, the gate capacitance of the K-OECT serves the same function.

The voltage to the gate of the Na-OECT is amplified with either inverting or non-inverting amplifier as discussed above and is a function of membrane voltage as $f(V_{mem}, t)$. The drain of the Na-OECT is applied with $E_{Na}$ while the source of the Na-OECT is connected to the node of $V_{mem}$. Two thresholds $V_{Na-m}$ and $V_{Na-s}$ are employed to simulate the activation and the inactivation of the Na channel, respectively. The current of Na channel is then written as a combination of two equations of OECTs operated in saturation region as:

$$i_{Na}(t) = \mu_{Na-m} C_V \frac{Wd}{L}[f(V_{mem}, t) - V_{mem} - V_{Na-m}]^2 \text{, } (V_{NA-m} < f(V_{mem}) < (V_{NA-m}+V_{NA-s})/2) \quad (5)$$

$$i_{Na}(t) = \mu_{Na-s} C_V \frac{Wd}{L}[f(V_{mem}, t) - E_{Na} - V_{Na-s}]^2 \text{, } ((V_{NA-m}+V_{NA-s})/2 < f(V_{mem}) < V_{NA-s}). \quad (6)$$

To further simulate the behaviors of c-OECNs, SPICE models based on equivalent circuits have been developed for both K-OECT and Na-OECT. The K-OECT SPICE model is a further update on the previously reported[1], with larger channel (W/L= 400/6 μm). The simulated transfer characteristics of K-OECT is presented in **Supplementary Figure 9a** along with the measured data. For the Na-OECT, a new equivalent circuit has been developed with a pair of PMOS (Ms) and NMOS (Mm) transistors, shown in **Supplementary Figure 8**. The simulated transfer characteristics of Na-OECT is shown in **Supplementary Figure 9b**.

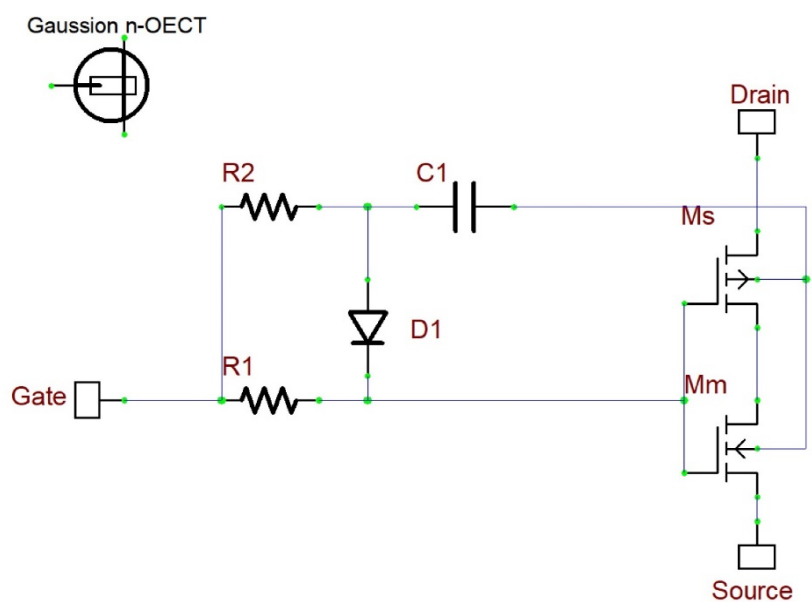

**Supplementary Figure 8**: Left upper: Symbol of the SPICE model for the BBL n-type OECTs with antiambipolar/Gaussion characteristics. The equivalent circuit consists of two resistors (R2=800 kΩ, R1=850 kΩ), one capacitor (C1=90 nF), one diode (D1), one PMOS (Ms) and one NMOS (Mm).

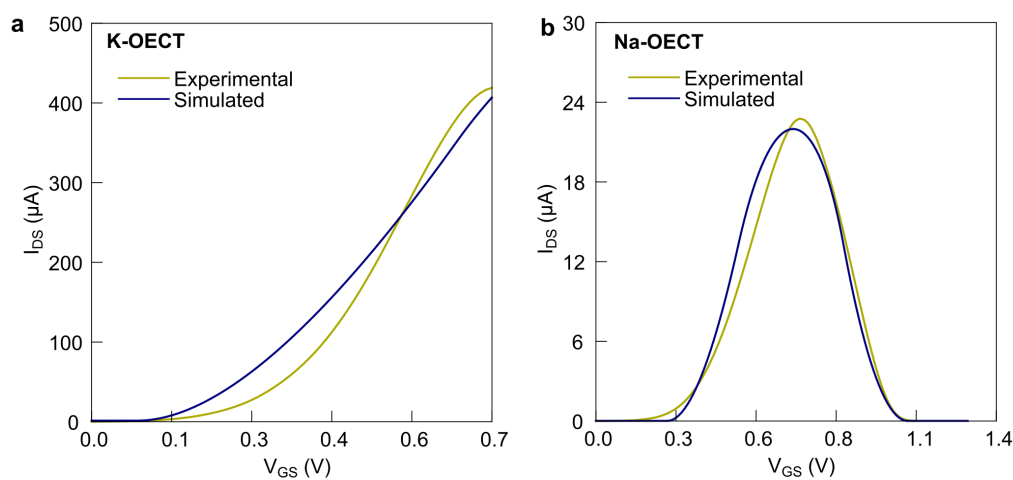

**Supplementary Figure 9:** Comparison of simulated and measured transfer characteristics of Na and K-OECTs.

With the SPICE models built, we simulate various spiking features of c-OECN and **Supplementary Figure 10** present a typical tonic spiking circuit. The simulated c-OECN spiking features are shown in **Supplementary Figure 11**.

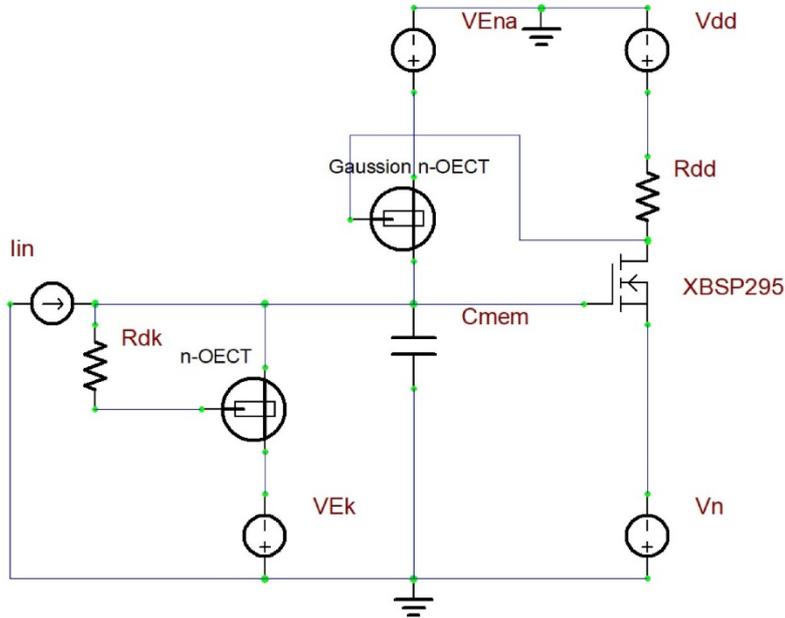

**Supplementary Figure 10**. A typical tonic spiking c-OECN with SPICE models.

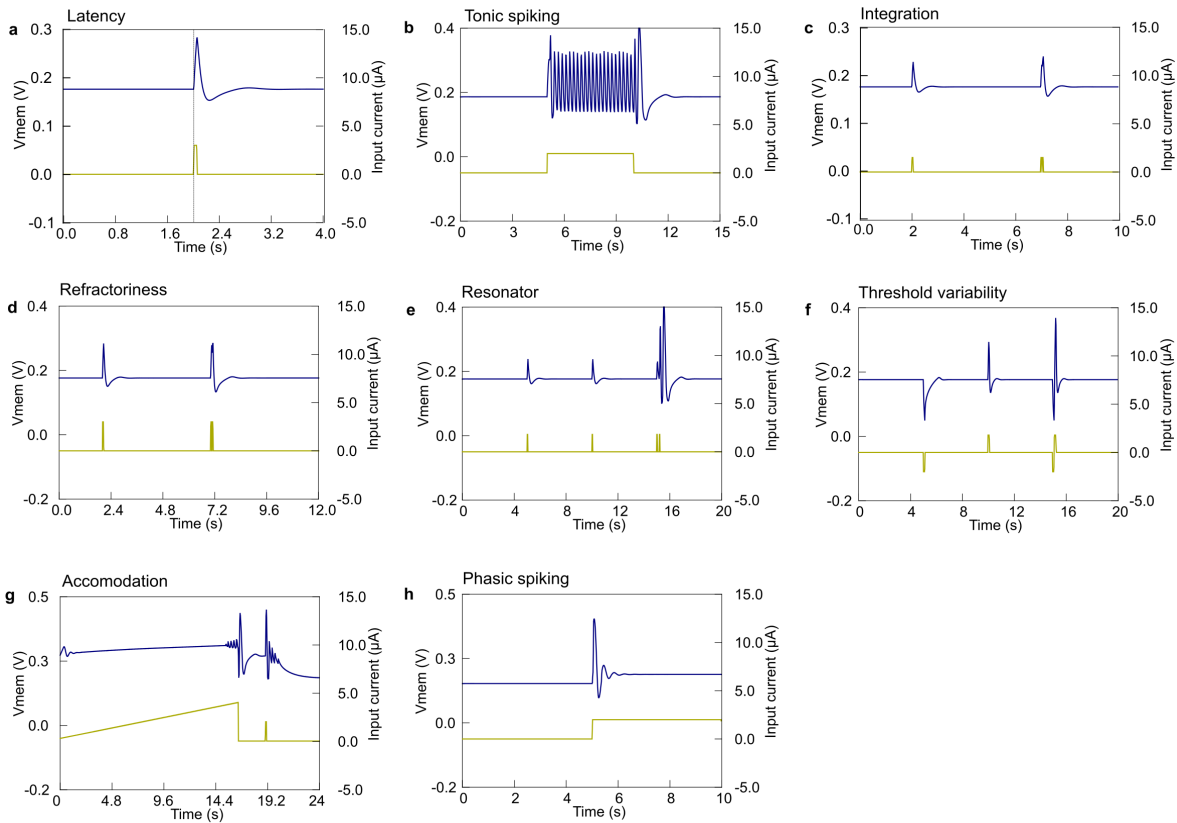

**Supplementary Figure 11**: Neural features based on SPICE simulations with K-OECT and Na-OECT models.

For comparison, we also investigate the possibility of using an NMOS transistor instead of K-OECT to represent the K channel, where one extra capacitor $C_{dk}$ is needed to adjust the delay of the activation of the K channel **(Supplementary Figure 12)**. It suggests that K-OECT can be replaced by other n-type enhacement-mode transistors (for example, NMOS, TFTs, etc.), however, the Na-OECT relies on the unique feature of the antiambipolar BBL OECTs. The experimental and simulated neural features of this circuit are shown in **Supplementary Figures 13 and 14**.

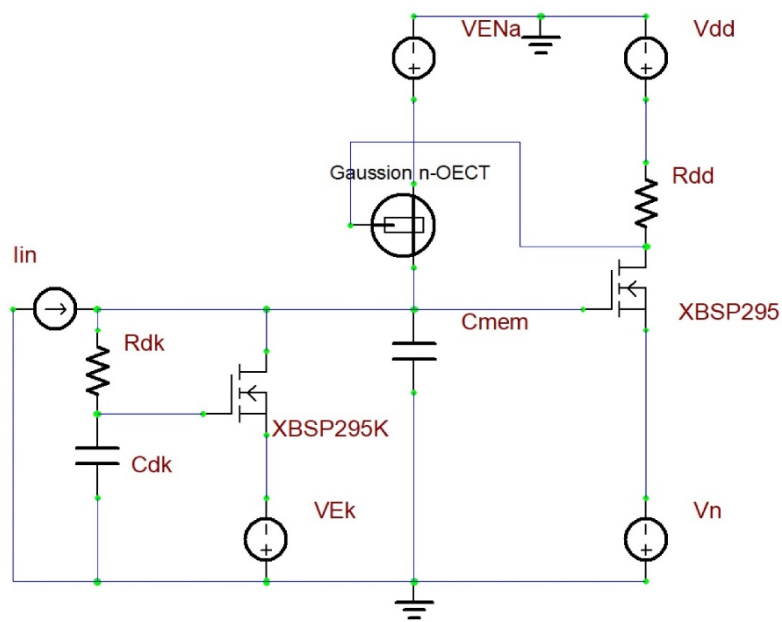

**Supplementary Figure 12**. A similar c-OECN as Supplementary Figure 11, with a NMOS transistor to replace the K-OECT.

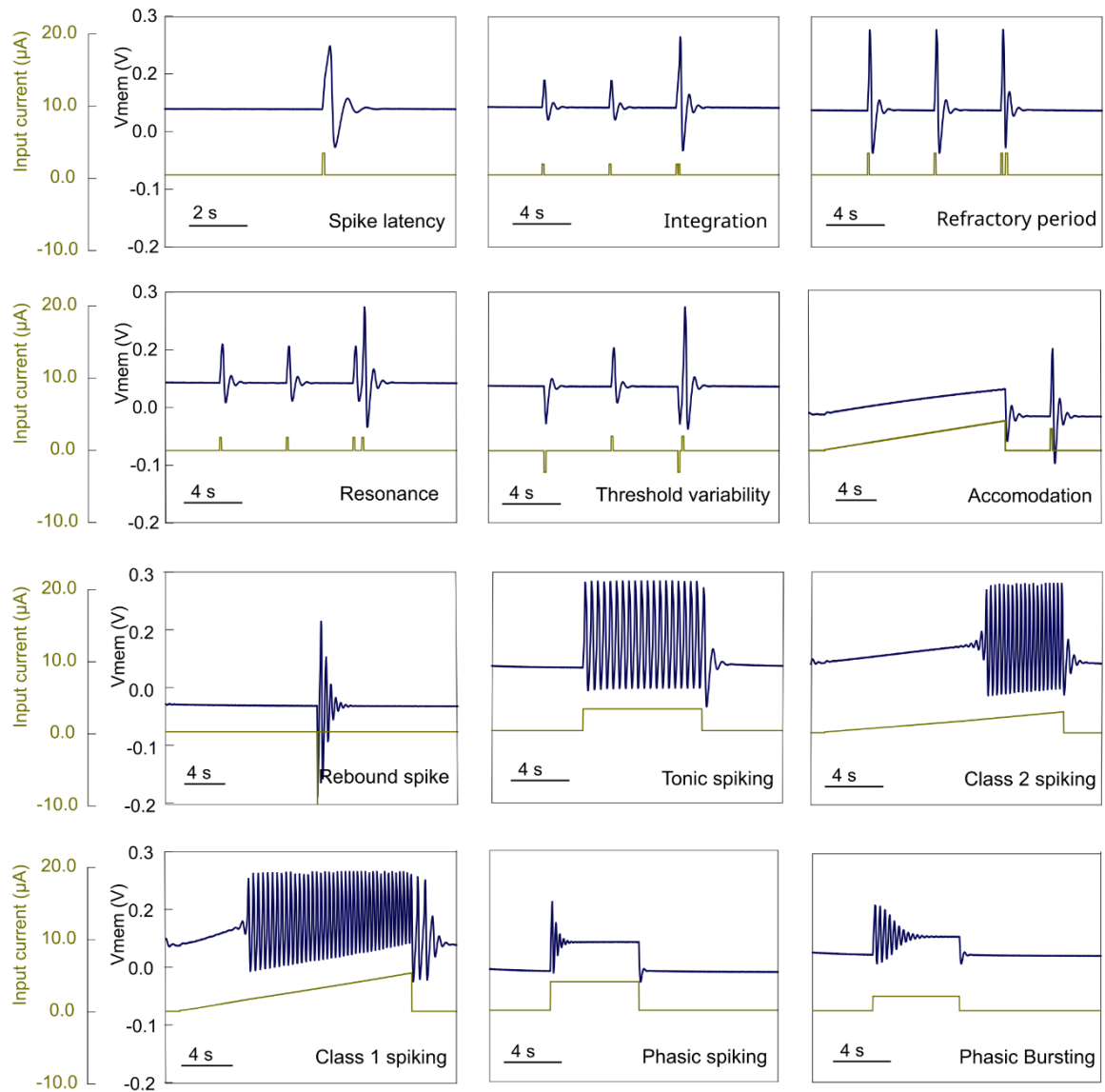

**Supplementary Figure 13 :** Features of the neuron using an NMOS transistor along with capacitor $C_{dk}$ instead of OECT for the K-channel.

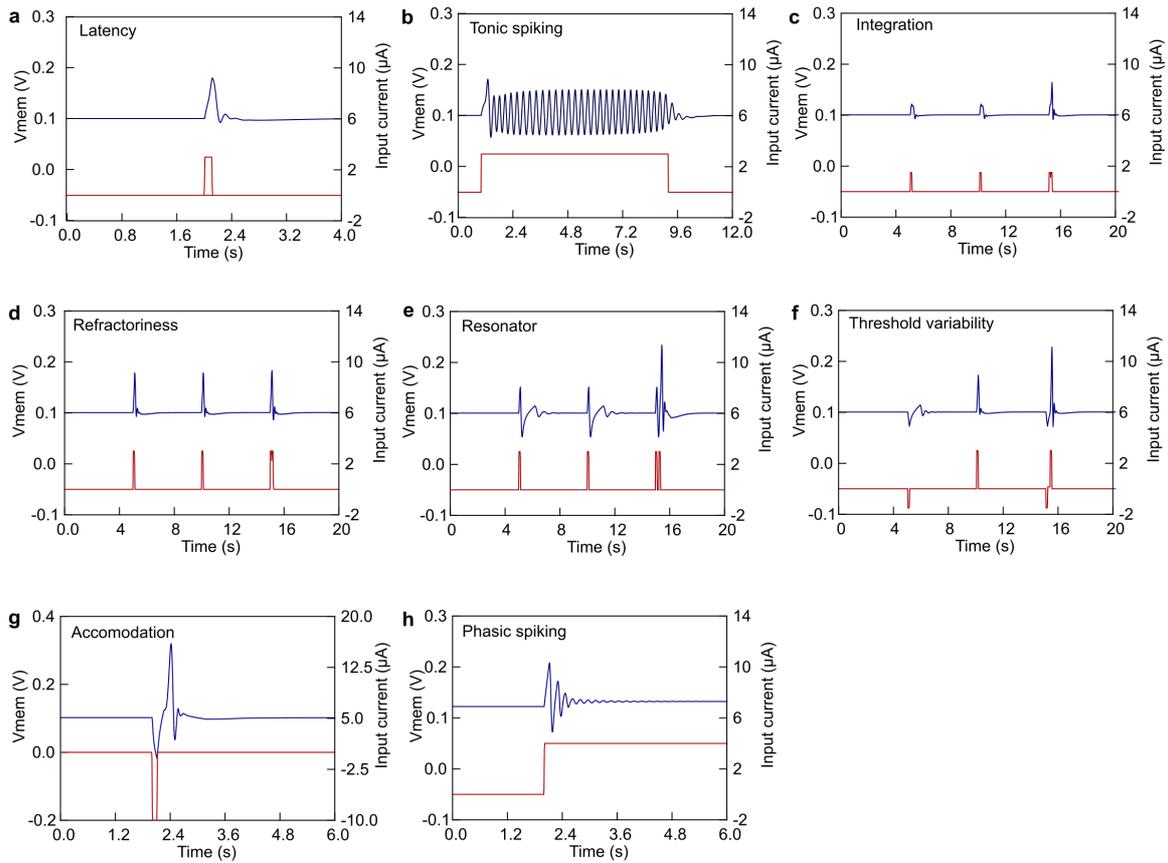

**Supplementary Figure 14 :** Simulated Features of the neuron using an NMOS transistor along with capacitor $C_{dk}$ instead of OECT for the K-channel.

**Supplementary Note 3 : Power consumption estimation**

The main components of power consumption in the circuit are the Na and K currents. **Supplementary Figure 15a** shows the sample action potential of a slow neuron with a period of 350 ms. The power consumption of the channels is calculated as

$$P_{channel} = (E_{Na}-V_{mem}) \times I_{Na} + (V_{mem} - E_K) \times I_K \qquad (7).$$

This gives a maximum power consumption of around 7 µW during the spike (**Supplementary Figure 15a**), excluding the inverter. For a comparison, conductance-based Si CMOS neurons consume around 15-60 µW power[3,4]. The energy consumed by this slow neural spiking is around 800 nJ /spike obtained by integrating the power over time (**Supplementary Figure 15b**). Since the same circuit has been used to realise neurons spiking at 80 Hz ( 12.5 ms time period), the energy consumption of such fast spiking is around 30 nJ /spike (excluding the inverter).

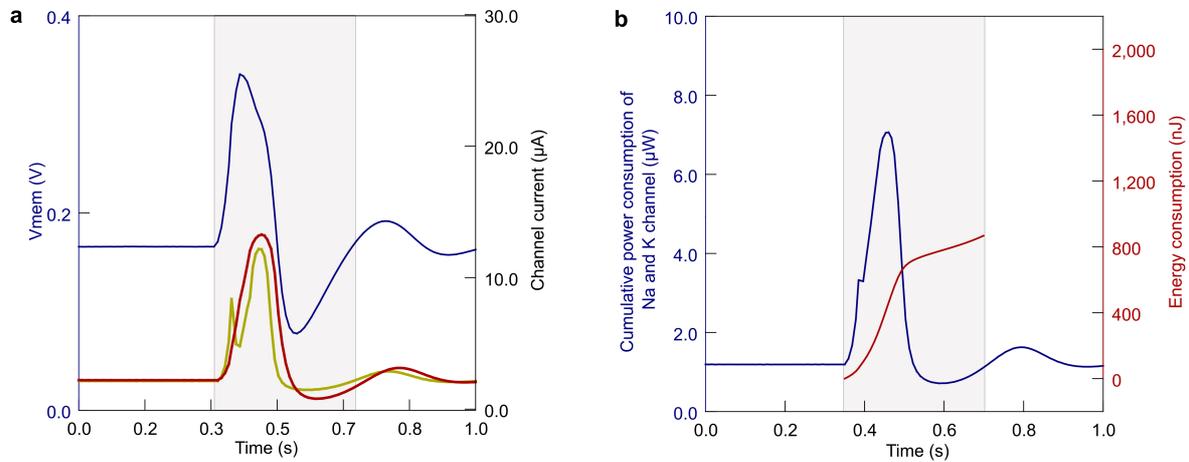

**Supplementary Figure 15 :** (a) Sample action potential of a slow neural spike (period of 350ms) showing the Na$^+$ and K$^+$ currents and (b) the corresponding power and energy consumption.

**Supplementary Note 4 : Features of neuron**

It is observed that there is delay between the applied input pulse and the generated spike in the c-OECN (**Figure 3d**). This latency enables encoding the input pulse strength because this delay will be proportional to input amplitude. The OECN can demonstrate integration of the input signals, ie when two inputs producing subthreshold spikes occur in a short interval of time, it is integrated to generate a spike and enable coincidence detection (**Figure 3e**). However, if a second input occurs during the hyperpolarisation phase of a previous action potential, there is no new spike generation, thus exhibiting refractoriness (**Figure 3f**). Due to the presence of subthreshold oscillations associated with action potential, the c-OECN exhibits resonance or frequency preference, ie when two subthreshold inputs occur at a frequency in phase with the oscillations, a spike is generated (**Figure 3g**). This enables frequency modulated interactions with the neuron. As opposed to the leaky integrate and fire-based OECN which exhibits a fixed threshold for spike generation, the c-OECN has a variable threshold that depends on the neuron's prior activity. As shown in **Figure 3h**, a signal which generates subthreshold output can create a spike if the threshold is lowered by a preceding inhibitory input. Rebound spike, where the c-OECN spikes in response to an inhibitory input (**Figure 3i**) and accommodation where a slowly varying input does not invoke spike while a smaller but sharper input (**Figure 3j**) can incite a spike can also be observed.

The c-OECN can emulate 3 classes of neurons by varying the threshold of the K-OECT (Figure 3k-m). For an $E_k$= -0.05 V, class 1 spiking is observed where the spiking frequency varies with the strength of the input. These neurons can encode the strength of input in frequency of spiking. Changing the $E_k$ to more negative values of -0.065V changes the behaviour to class 2 spiking where the c-OECN spikes at high amplitude inputs without much modulation of frequency. An even larger negative $E_k$ causes Phasic bursting and spiking which is commonly refereed as class 3 spiking where spikes or bursts are generated at the onset of the input helping the encode the beginning of a stimulation. Bursting helps reduce synaptic failure, transmit input saliency, and enable selective communication between neurons.

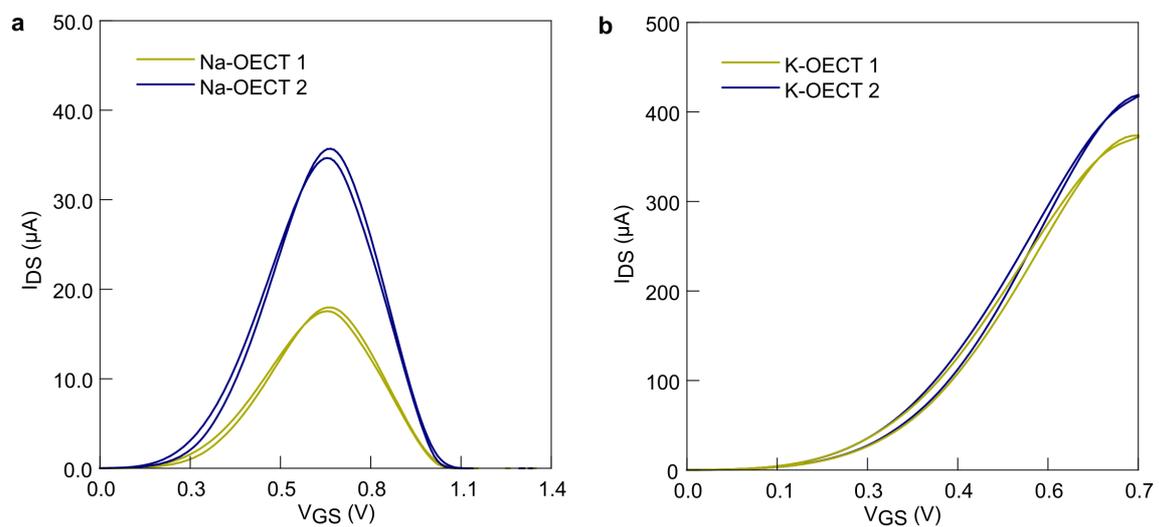

**Supplementary Figure 16 :** The transfer characteristics of the Na-OECTs and K-OECTs used in the circuit to to obtain various neural features in Supplementary Table 1.

**Supplementary Table 1:** Parameters used for different configuration of c-OECN.

| | Feature | Na-OECT | K-OECT | $C_{mem}$ (μF) | $R_{dk}$ (kΩ) | $E_k$ (V) | $E_{Na}$ (V) | $V_{DD}$ (V) | $E_N$ (V) | $R_{DD}$ (kΩ) |
|---|---|---|---|---|---|---|---|---|---|---|
| 1 | Tonic spiking | Na-OECT 1 in Supplementary Figure 17 | K-OECT 1 in Supplementary Figure 17 | 1 | 470 | -0.05 | 0.5 | 1.25 | -0.66 | 12 |
| 2 | Latency | | | | | | | | | |
| 3 | Subthreshold oscillations | | | | | | | | | |
| 4 | Integration | | | | | | | | | |
| 5 | Refractoriness | | | | | | | | | |
| 6 | Resonance | | | | | | | | | |
| 7 | Variable Threshold | | | | | | | | | |
| 8 | Rebound spike | | | | | | | | | |
| 9 | Accomodation | | | | | | | | | |
| 10 | Phasic Bursting | | | | | -0.07 | | | | |
| 11 | Phasic spiking | | | | | -0.072 | | | | |
| 12 | Class 1 spiking | | | | | -0.05 | | | | |
| 13 | Class 2 spiking | | | | | -0.065 | | | | |
| 15 | Calcium control | | | | | -0.05 | | | | |
| 16 | GABA control | Na-OECT 2 in Supplementary Figure 17 | K-OECT 2 in Supplementary Figure 17 | | | -0.012 | | | -0.65 | |
| 14 | Stochastic spiking | | | | | 0.025 | | | | |
| 17 | Fast neuron (45Hz) | | | - | 68 | 0.01 | | | | |
| 18 | Fast neuron (80Hz) | | | - | 27 | 0.02 | | | | |